\documentclass{patmorin}
\usepackage{pat}
\usepackage[T1]{fontenc}
\usepackage[utf8]{inputenc}
\usepackage{mathtools}
\usepackage{thmtools}
\usepackage{thm-restate}

\usepackage[inline]{enumitem}
\usepackage{soul}
\usepackage[normalem]{ulem}

\crefname{p}{}{}
\creflabelformat{p}{#2(#1)#3}

\usepackage{todonotes}

\usepackage[longnamesfirst,numbers,sort&compress]{natbib}

\DeclarePairedDelimiter{\abs}{\lvert}{\rvert}

\title{\MakeUppercase{Containment Problems in Convex Sets}%
  \thanks{This research was partly funded by NSERC.}}

\author{Prosenjit Bose%
  \thanks{School of Computer Science, Carleton University.},\quad
  Bobby Miraftab\footnotemark[2],\quad
  and Tyler Tuttle\footnotemark[2]}

\date{}

\begin{document}

\maketitle

\begin{abstract}
Let $P$ be a set of $n$ points in the plane, and let $C$ be a compact, convex subset of the plane that contains the origin in its interior. We study the problem of finding two points $p$ and $q$ of $P$ with the property that any homothet of $C$ that contains $p$ and $q$ must contain at least a constant fraction of the points of $P$. We give proofs for several cases. When $C$ is strictly convex we achieve a fraction of $1/5$, and when $C$ is a general convex region we achieve a fraction of $1/6e$. We show how the ideas of this second bound can be used to prove a lower bound for a version of the problem in $\mathbb{R}^d$ when $C$ is a $d$-ball. We also consider a few special cases. When $C$ is a triangle we get a lower bound of $1/12$, and when we allow that triangle to be rotated by $\pi$ radians in addition to being scaled and translated we get a lower bound of $1/768$. In proving this last result, we improve the state of the art on the version of this problem in three dimensions, where $C$ is a parallelepiped. We also consider the algorithmic aspects of finding such a pair of points.
\end{abstract}

\paragraph{AI Declaration.}
We did not use generative AI tools in the creation of this paper. All research, writing, and revisions are our own work.

\section{Introduction}

Let $P$ be a set of $n$ points in the plane in general position. In 1988, Neumann-Lara and Urrutia~\cite{Neumann-LaraU88} proved that there exist two points $p$ and $q$ in $S$ such that any disk that contains $p$ and $q$ must contain at least $\lceil(n-2)/60 \rceil$ points of $S$. A series of papers followed that improved this result~\cite{Hayward89,HaywardRW89,BaranySSU89}, with Edelsbrunner et al.~\cite{EdelsbrunnerHSX89} giving a lower bound of approximately $n/4.7$ the following year. Hayward et al.~\cite{HaywardRW89} constructed a set of points $P$ such that for any two points $p$ and $q$ in $P$, there exists a disk that contains $p$, $q$, and at most $n/4$ other points of $P$.

Several variations on this problem have been studied in the literature. Hayward et al.~\cite{HaywardRW89} considered restricting $P$ to be in convex position. Akiyama et al.~\cite{AkiyamaIUU96} considered the problem of finding a pair of points $p$ and $q$ such that the disk with $pq$ as a diameter contains many points.
Additionally, a coloured version of the problem has been studied~\cite{Valencia05,DBLP:journals/dcg/Prodromou07,ClaverolHM21}, as well as a generalization to higher dimensions~\cite{BaranySSU89,SmorodinskySW08}.

Extensions of the problem to shapes other than disks have been studied as well. Ishigami~\cite{Ishigami95} showed that there exist two points $p$ and $q$ such that the smallest axis-aligned box that contains $p$ and $q$ must contain at least $cn$ points. Barany et al.~\cite{BaranyL90} considered replacing the notion of a disk with that of an ellipse, or more generally a quadric. Abellanas et al.~\cite{AbellanasHKNU97} considered a very general version of the problem, with arbitrary convex sets in place of both the disk and the set of points.

Finally, Bose et al.~\cite{BoseEOST25} considered this problem in the setting of a simple polygon, with a geodesic disk replacing the notion of a disk. Bose et al.~\cite{BoseET25} also gave algorithms for several versions of the problem, both for disks in the plane and geodesic disks in a simple polygon.

Recall that a subset $C$ of $\mathbb{R}^d$ is called \emph{convex} if, for any two points $p$ and $q$ in $C$, the line segment $pq$ is contained in $C$. The subset $C$ is called \emph{strictly convex} if the relative interior of $pq$ is contained in the interior of $C$. Equivalently, $C$ is strictly convex if no three points on the boundary of $C$ are collinear.

Let $C$ be a compact, convex subset of $\mathbb{R}^d$ that contains the origin in its interior. A homothet of $C$ is any subset of $\mathbb{R}^d$ that can be obtained by uniformly scaling and then translating $C$. In other words, any set of the form $\lambda C + p = \{\lambda q + p : q \in C\}$.

In this paper we consider the following version of the problem. Let $P$ be a set of $n$ points in the plane, and let $C$ be a compact, convex subset of the plane that contains the origin in its interior. We show that there exist two points $p$ and $q$ in $P$ such that any homothet of $C$ that contains both $p$ and $q$ must contain $cn$ points, for some constant $c$. We also show that there exist two points $p$ and $q$ in $P$ such that any homothet of $C$ that has $p$ and $q$ on its boundary contains between $c'n$ and $(1-c')n$ other points of $P$, for a different constant $c'$. We give two different lower bounds for the general case in the plane, one assuming that $C$ is strictly convex and one that does not. We also consider the special case where $C$ is a triangle in the plane, and the case where homothets of the triangle are also allowed to be scaled by negative numbers (equivalently, rotated by $\pi$ radians). In proving this last result, we improve the state of the art for boxes in $\mathbb{R}^3$.

Finally, we consider the algorithmic aspects of these variations of the problem. For general convex sets, we discuss properties that are necessary for efficient algorithms to exist. For triangles in $\mathbb{R}^2$ and boxes in $\mathbb{R}^3$, we give algorithms that can efficiently find pairs with the given property. All of the algorithms we give are randomized.

\section{Homothets of a convex set}

In this section we give two bounds for when $C$ is a convex subset of the plane. The first bound assumes that $C$ is strictly convex, while the second one does not. This additional assumption allows us to prove a better lower bound. First, we introduce some notions that will be useful throughout the section.

Fix a compact, convex set $C$ in $\mathbb{R}^2$ that contains the origin in its interior. The \emph{convex distance function} associated with $C$ is the function $d_C : \mathbb{R}^2 \times \mathbb{R}^2 \to \mathbb{R}$ defined by
\begin{equation}
    d_C(p,q) = \min\{\lambda \ge 0 : q \in \lambda C + p\}.
\end{equation}
Convex distance functions satisfy certain properties that will be useful to us. It is straightforward to show that $d_C(p,q) > 0$ if $p \ne q$, and $d_C(p,q) = 0$ if $p=q$. Chew and Drysdale proved that $d_C$ satisfies the triangle inequality~\cite{DBLP:conf/compgeom/ChewD85}. In general, we do not have $d_C(p,q) = d_C(q,p)$. However, if $C$ is symmetric about the origin (meaning $p \in C$ if and only if $-p \in C$), then $d_C$ is symmetric and hence a metric on $\mathbb{R}^2$.

Certain well-known metrics are in fact convex distance functions. The standard Euclidean metric corresponds to the case where $C$ is a disk. In fact, all $L_p$ metrics are convex distance functions, with the unit ball $\{(x,y) \in \mathbb{R}^2 : x^p + y^p \le 1\}$ as the convex region $C$.

Given two points $p$ and $q$, define the bisector $b_C(p,q)$ to be the set
\begin{equation}
    \{r \in \mathbb{R}^2 : d_C(p,r) = d_C(q,r)\}.
\end{equation}
If $C$ is strictly convex, then $b_C(p,q)$ is homeomorphic to a line for all $p$ and $q$. If $C$ is not strictly convex, then $b_C(p,q)$ may contain regions with positive area.

Let $H_C(p,q)$ be the set of points which are closer to $p$ than $q$, with respect to $d_C$. That is,
\begin{equation}
    H_C(p,q) = \{r \in \mathbb{R}^2 : d_C(p,r) < d_C(q,r)\}.
\end{equation}
The three sets $H_C(p,q)$, $H_C(q,p)$, and $b_C(p,q)$ partition the plane, and $b_C(p,q)$ is the boundary of both $H_C(p,q)$ and $H_C(q,p)$. By analogy to the case where $d_C$ is the Euclidean metric, we will call this the ``halfplane'' bounded by $b_C(p,q)$ that contains $p$.

\subsection{Strictly convex sets}

Throughout this section we assume that $C$ is a strictly convex set. The fact that $C$ is strictly convex implies that the bisector of two points is homeomorphic to a line, and that the bisector of \emph{three} points is either empty if the three points are collinear, or a single point if they are not. This second fact implies that there is a unique homothet of $C$ that contains three given points on its boundary.

Let $P$ be a set of $n$ points in $\mathbb{R}^2$, and let $C$ be a strictly convex, compact subset of $\mathbb{R}^2$ that contains the origin in its interior. Let $Q$ be a subset of $P$ of size $k$. The order-$k$ Voronoi region of $Q$, denoted $\mathrm{VR}_k(Q)$, is the set of points in $\mathbb{R}^2$ whose $k$ nearest neighbours are the set $Q$, when measured using the convex distance function $d_C$. Formally, we can define it to be the set
\begin{equation}
    \mathrm{VR}_k(Q) = \big\{r \in \mathbb{R}^2 : \max\{d_C(r,q) : q \in Q\} < \min\{d_C(r,p) : p \in P \setminus Q\}\big\}.
\end{equation}
There is another, equivalent way to define the Voronoi regions, in terms of halfplanes $H_C(p,q)$:
\begin{equation}
    \mathrm{VR}_k(Q) = \bigcap_{p \in P\setminus Q, q \in Q} H_C(q,p).
\end{equation}
The boundary of a Voronoi region is composed of pieces of bisectors. This is a generalization of the standard Euclidean higher-order Voronoi diagrams.

The bisectors of a convex distance function satisfy the axioms of an abstract Voronoi diagram, as defined by~\cite{DBLP:journals/comgeo/KleinLN09}. An abstract Voronoi diagram (AVD) consists of a set of curves (which in our case are the bisectors defined by pairs of points) that satisfy a certain set of axioms that allow one to define structures similar to Voronoi diagrams. In particular, higher-order abstract Voronoi diagrams were studied by Abellanas et al.~\cite{AbellanasHKNU97}, and they proved certain properties that will be of use to us.

We note that proving convex distance functions satisfy the axioms of an abstract Voronoi diagram is nontrivial. The original set of axioms included the assumption that any two bisectors intersect at most a finite number of times. However, it was shown that there exist strictly convex regions, even with smooth boundaries, whose convex distance functions admit bisectors that can intersect infinitely many times~\cite{DBLP:journals/ijcga/CorbalanMR96}. Partly because of this, Klein et al.~\cite{DBLP:journals/comgeo/KleinLN09} removed that assumption from their list of axioms, and proved all the known properties of abstract Voronoi diagrams without relying on it.

There is one important technical detail to consider. In the case where $C$ is a circle, the centre of any disk with two given points $p$ and $q$ on the boundary must lie on the bisector of $p$ and $q$. This is not true in general. The bisector is the set of points $r$ such that $d_C(p,r) = d_C(q,r)$, but what we want is a point $r$ such that $d_C(r,p) = d_C(r,q)$. This can be solved by considering the Voronoi diagrams defined by the convex distance function $d_{-C}$, where $-C = \{-x \in \mathbb{R}^2 : x \in C\}$. For any two points in the plane, we have $d_C(p,q) = d_{-C}(q,p)$. So if $\lambda C + r$ is a homothet of $C$ such that $p$ and $q$ are both on the boundary of $\lambda C + r$, then $r$ must lie on $b_{-C}(p,q)$.

\begin{thm}
\label{thm:strictly-convex-sets}
Let $C$ be a compact, strictly convex subset of $\mathbb{R}^2$ that contains the origin in its interior, and let $P$ be a set of $n$ points in $\mathbb{R}^2$. There exists a pair of points $\{p,q\}$ in $P$ such that any homothet of $C$ that contains $\{p,q\}$ contains at least $n/5$ other points of $P$.
\end{thm}

\begin{proof}
Consider two points $p$ and $q$ of $P$, and the bisector $b_{-C}(p,q)$. Since $C$ is strictly convex, so is $-C$, and $b_{-C}(p,q)$ is homeomorphic to a line.
Imagine continuously transforming a homothet of $C$ so that its centre moves along $b_{-C}(p,q)$, and the points $p$ and $q$ remain on the boundary.
There are $n-2$ homothets of $C$ that have $p$, $q$ and some third point of $P$ on its boundary. The centre of each of these homothets is a point on $b_{-C}(p,q)$, and these $n-2$ centres partition $b_{-C}(p,q)$ into $n-1$ pieces, two of which are unbounded. We will refer to these pieces as the segments of $b_{-C}(p,q)$. Any homothet with $p$ and $q$ on its boundary whose centre is on some segment $s$ contains the exact same set of points of $P$ in its interior. Let $\rho(s)$ denote the size of this set, which we will call the weight of $s$. The edges of the order-$k$ Voronoi diagram are precisely the segments with weight $k-1$.

If some bisector $b_{-C}(p,q)$ contains only edges of large weight, then every homothet of $C$ that contains both $p$ and $q$ must contain many points. To do this, consider the union of the edges of the order-$1$ through order-$k$ Voronoi diagrams of $P$. Along one bisector, these edges form some number of \emph{connected components}: maximal connected subsets of $b_{-C}(p,q)$. Let $\lambda_k$ be the total number of connected components across all bisectors. If $\lambda_k < \binom{n}{2}$, then there must exist a bisector $b_C(p,q)$ with no piece of weight less than $k$, and therefore every homothet of $C$ that contains $p$ and $q$ contains at least $k$ other points.

As noted earlier, the system of bisectors induced by a convex distance function satisfy the axioms of an abstract Voronoi diagram. Bose et al.~\cite{BoseEOST25} proved that, for abstract Voronoi diagrams, the bound
\begin{equation}
    \lambda_k \le 3kn - \frac52k^2 - \frac52k
\end{equation}
holds. We note that their proof is for a specific type of abstract Voronoi diagram, but their proof only uses properties of higher-order abstract Voronoi diagrams and so it implies the same bound for any AVD.

As long as $3kn - \frac52k^2 -\frac52k < \binom{n}{2}$, we are done. This inequality is satisfied as long as $k < n/5$.
\end{proof}

Using similar techniques, Bose et al.\ proved the existence of a pair of points such that any homothet of $C$ that contains those two points on its boundary must contain many points, but also not contain too many points. This proof also carries over to the general setting where $C$ is a strictly convex region.

\begin{cor}
\label{thm:strictly-convex-inside-outside}
Let $C$ be a compact, strictly convex subset of $\mathbb{R}^2$ that contains the origin in its interior, and let $P$ be a set of $n$ points in $\mathbb{R}^2$. There exists a pair of points $\{p,q\}$ in $P$ such that any homothet of $C$ with $p$ and $q$ on its boundary contains at least $n/13.08$, and at most $(1 - 1/13.08)n$, other points of $P$.
\end{cor}

\begin{proof}
In order to find homothets of $C$ that do not contain too many points, we need to find a bisector $b_{-C}(p,q)$ with no segments of high weight. Similar to the proof of \Cref{thm:strictly-convex-sets}, consider the union of the order-$(n-1)$ through order-$(n-k)$ Voronoi diagrams of $P$. Let $\overline\lambda_{n-k}$ be the sum of the number of connected components over all bisectors of this union. If $\lambda_k + \overline\lambda_{n-k} < \binom{n}{2}$, then there exists at least one bisector $b_{-C}(p,q)$ with no segment of weight less than $k$ or greater than $n-k$. This means that every homothet of $C$ that has $p$ and $q$ on its boundary will contain at least $k$ points and at most $n-k$ points of $P$.

Bose et al.~\cite{BoseEOST25} proved that, for abstract Voronoi diagrams, the bound
\begin{equation}
    \lambda_k + \overline\lambda_{n-k} \le 7kn - 6k^2 - 6k
\end{equation}
holds. Once again we note that their proof holds for any AVD. We have $7kn - 6k^2 - 6k < \binom{n}{2}$ as long as
\begin{equation}
    k < \frac{7n - 6 - \sqrt{37n^2 - 72n + 36}}{12} = \bigg(\frac{7 - \sqrt{37}}{12}\bigg)n + O(1) \approx \frac{1}{13.08}n.
\end{equation}
\end{proof}

\subsection{General convex sets}
\label{sec:general-convex}

In the previous section we relied heavily on the fact that $C$ is strictly convex, because that implied that bisectors were homeomorphic to lines, which is what in turn allowed us to use the properties of abstract Voronoi diagrams. In this section, we provide a different proof that does not require $C$ to be strictly convex, at the cost of a worse lower bound.

We say that the line segment $pq$ has order $k$ if there exists a homothet of $C$ with $p$ and $q$ on the boundary that contains $k$ other points of $P$. The order-$k$ Delaunay graph $D_k(P)$ consists of all edges of order at most $k$. Note that if $pq$ is \emph{not} an edge of the order-$k$ Delaunay graph, then every homothet of $C$ with $p$ and $q$ on the boundary contains more than $k$ other points of $P$. In other words, every segment of $b_{-C}(p,q)$ has weight greater than $k$. Therefore, we want to compute the highest value of $k$ for which the order-$k$ Delaunay graph is not the complete graph.

We note that, if $C$ is strictly convex, then the quantity $\lambda_k$ defined in the previous section is an upper bound on the number of edges of the order-$k$ Delaunay graph. We present an alternative way to bound the number of edges that does not rely on the assumption that $C$ is strictly convex, based on the technique of Clarkson and Shor.

\begin{thm}
\label{thm:convex-sets}
Let $C$ be a compact, convex subset of $\mathbb{R}^2$ that contains the origin in its interior, and let $P$ be a set of $n$ points in $\mathbb{R}^2$. There exists a pair of points $\{p,q\}$ in $P$ such that any homothet of $C$ that contains $\{p,q\}$ contains at least $(n+1)/6e - 1 \approx n/16.31$ other points of $P$.
\end{thm}

\begin{proof}
Choose a random sample $R$ of $P$, with each point of $P$ included independently with probability $\alpha = (k+1)^{-1}$. If $pq$ is an edge of the order-$k$ Delaunay graph of $P$, then there exists a homothet $\lambda C + x$ that contains $p$ and $q$ on it's boundary and at most $k$ other points of $P$. With probability at least $\alpha^2(1-\alpha)^k$, $p$ and $q$ are in $R$ and no other point in $P \cap (\lambda C + x)$ is in $R$. This implies that $pq$ is an edge of the order-$0$ Delaunay graph of $R$ with probability at least $\alpha^2(1-\alpha)^k$, and so we can bound the expected number of edges of $D_0(R)$ as
\begin{equation}
    E[\lvert D_0(R) \rvert] \ge \alpha^2(1-\alpha)^k\lvert D_k(P) \rvert.
\end{equation}
On the other hand, it is known that the order-$0$ Delaunay graph is plane~\cite{DBLP:journals/jocg/BoseCCS10}, and so it has at most $3\lvert R \rvert - 6$ edges, so $E[\lvert D_0(R) \rvert] \le 3E[\lvert R \rvert] - 6 = 3\alpha n - 6$. Therefore
\begin{equation}
    \lvert D_k(P) \rvert \le \frac{3\alpha n - 6}{\alpha^2(1-\alpha)^k} < 3en(k+1)
\end{equation}
where the final inequality follows from the fact that $1 - x < e^{-x}$ for all $x > 0$.

As long as $3en(k+1) < \binom{n}{2}$, then $D_k(P)$ is not the complete graph. This implies that there exists a pair of vertices $p, q$ in $D_k(P)$ that are non-adjacent, which means that there does not exist any homothet $\lambda C + x$ containing at most $k$ points of $P$ in its interior with $p, q$ on its boundary. Thus, this pair has the desired property. Rearranging the inequality $3en(k+1) < \binom{n}{2}$ yields the claimed bound of $k < (n+1)/6e - 1$.
\end{proof}

We can also prove that there exists two points $p$ and $q$ with the property that any homothet of $C$ with $p$ and $q$ on its boundary must contain at least $cn$ and at most $(1-c)n$ points of $P$, where $c$ is a constant. To do this we will define the farthest-point Delaunay graph of $P$. We say that $pq$ is an edge of the order-$k$ farthest-point Delaunay graph if there exists a homothet of $C$ with $p$ and $q$ on its boundary that contains all but $k$ points of $P$ in its interior. Let $\overline{D}_k(P)$ denote the order-$k$ farthest-point Delaunay graph of $P$.

\begin{cor}
\label{cor:convex-sets-inside-outside}
Let $C$ be a compact, convex subset of $\mathbb{R}^2$ that contains the origin in its interior, and let $P$ be a set of $n$ points in $\mathbb{R}^2$. There exists a pair of points $\{p,q\}$ in $P$ such that any homothet of $C$ with $p$ and $q$ on its boundary contains at least $n/27.18$, and at most $(1-1/27.18)n$, other points of $P$.
\end{cor}

\begin{proof}
Choose a random sample $R$ of $P$, where each point is included with probability $\alpha = (k+1)^{-1}$. We can bound $\abs{\overline{D}_k}$ using a similar argument to \Cref{thm:convex-sets}. Let $pq$ be an edge of the order-$k$ farthest-point Delaunay graph of $P$. Then with probability at least $\alpha^2(1-\alpha)^k$, the edge $pq$ is in the order-$0$ farthest-point Delaunay graph of $R$. It is also known that the order-$0$ farthest-point Delaunay graph of $R$ is an outerplanar graph (because the farthest-point Voronoi diagram of an AVD is a tree~\cite{DBLP:journals/comgeo/BohlerCKLPZ15}), and so it has at most $2\abs{R}-3$ edges. Therefore we have
\begin{equation}
    \abs{\overline{D}_k(P)} \le \frac{2\alpha n - 3}{\alpha^2(1-\alpha)^k} < 2en(k+1).
\end{equation}
If we add this to our bound for $\abs{D_k(P)}$, we get
\begin{equation}
    \abs{D_k(P)} + \abs{\overline{D}_k(P)} < 5en(k+1).
\end{equation}
Thus if $5en(k+1) < \binom{n}{2}$, there must exist a pair of points $p$ and $q$ such that $pq$ is neither an edge of $D_k(P)$ nor $\overline{D}_k(P)$. This is true for $k < (n+1)/10e - 1 \approx n/27.18$.
\end{proof}

\section{Homothets of a ball in higher dimensions}

In this section we show how to generalize the argument of \Cref{thm:convex-sets} to more than two dimensions, in the special case where $C$ is a $d$-ball. Instead of looking for a pair of points, we will look for a subset of $d$ points such that any ball that contains those $d$ points will contain a constant fraction of the points of $P$. We can define the order-$k$ Delaunay graph in $\mathbb{R}^d$. Let $P$ be a set of $n$ points in $\mathbb{R}^d$. We say that a subset of $d+1$ points of $P$ form a simplex of the order-$k$ Delaunay graph if the $d$-ball with those $d+1$ points on the boundary contains at most $k$ points of $P$ in its interior.

\begin{thm}
\label{thm:d-balls}
Let $P$ be a set of $n$ points in $\mathbb{R}^d$, where $d \ge 3$. There exists a subset $Q$ of $P$ of size $d$ such that any $d$-ball that contains $Q$ also contains at least
\begin{equation*}
    \bigg(\frac{\ceil{d/2}!}{2^{\ceil{d/2}+1}e(d+1)d^d}\bigg(\frac{\ln\ceil{d/2}}{\ceil{d/2}}\bigg)^{\ceil{d/2}}\bigg)^{\frac{1}{\floor{d/2}}}n - 1
\end{equation*}
other points of $P$ in its interior.
\end{thm}

\begin{proof}
The proof follows the same structure as \Cref{thm:convex-sets}. Choose a random sample $R$ of $P$, with each point of $P$ included independently with probability $\alpha = (k+1)^{-1}$. Let $T_k(P)$ denote the number of $(d-1)$-simplices in the order-$k$ Delaunay graph of $P$. If $S$ is a $(d-1)$-simplex of the order-$k$ Delaunay graph of $P$, then there exists a $d$-ball $B$ with the vertices of $S$ on its boundary that contains at most $k$ other points of $P$ in its interior. With probability at least $\alpha^d(1-\alpha)^k$, the vertices of $S$ are in $R$ and no other point of $P \cap B$ is in $R$. This implies that $S$ is a $(d-1)$-simplex of the order-$0$ Delaunay graph of $R$ with probability at least $\alpha^d(1-\alpha)^k$, and so we get
\begin{equation}
    E[T_0(R)] \ge \alpha^d(1-\alpha)^k T_k(P).
\end{equation}

Since every $d$-simplex of a Delaunay graph is bounded by $d+1$ many $(d-1)$-simplices, the number of $(d-1)$-simplices is at most $d+1$ times the number of $d$-simplices. The number of $d$-simplices of the order-$0$ Delaunay graph is related to the number of facets of the convex hull of a set of points in $\mathbb{R}^{d+1}$. By mapping a point $(x_1, x_2, \dots, x_d)$ to the point $(x_1, x_2, \dots, x_d, x_1^2 + x_2^2 + \cdots + x_d^2)$, the $d$-simplices of the order-$0$ Delaunay graph are in bijection with the facets of the lower hull of the image of $P$. Thus the total number of facets of the convex hull is an upper bound on the number of $d$-simplices in the order-$0$ Delaunay graph of $P$. The Upper Bound Theorem~\cite{mcmullen1970maximum,ziegler2012lectures} allows us to bound the number of facets of a polytope in $\mathbb{R}^{d+1}$. In particular, this theorem implies that the number of facets of any $(d+1)$-polytope is at most
\begin{equation}
    \binom{\abs{R} - \ceil{(d+1)/2}}{\ceil{d/2}} + \binom{\abs{R} - \ceil{d/2} - 1}{\floor{d/2}} < 2\binom{\abs{R}}{\ceil{d/2}} \le \frac{2}{\ceil{d/2}!}\abs{R}^{\ceil{d/2}}.
\end{equation}
Therefore, $T_0(R) \le \big(2(d+1)/\ceil{d/2}!\big)\abs{R}^{\ceil{d/2}}$.

Since $\abs{R}$ is a binomial random variable, a result of Lata{\l}a~\cite[Corollary 3]{latala1997estimation} implies
\begin{equation}
    E[\abs{R}^{\ceil{d/2}}] \le \bigg(\frac{2\ceil{d/2}}{\ln\ceil{d/2}}\bigg)^{\ceil{d/2}}E[\abs{R}]^{\ceil{d/2}}
\end{equation}
so
\begin{equation}
    E[T_0(R)]
    \le E\bigg[\frac{2(d+1)}{\ceil{d/2}!}\abs{R}^{\ceil{d/2}}\bigg]
    \le \frac{2(d+1)}{\ceil{d/2}!} \bigg(\frac{2\ceil{d/2}}{\ln\ceil{d/2}}\bigg)^{\ceil{d/2}}(\alpha n)^{\ceil{d/2}}.
\end{equation}

Combining this with our lower bound for $E[T_0(R)]$, we get
\begin{align*}
    T_k(P)
    &\le \frac{2(d+1)}{\alpha^d(1-\alpha)^k\ceil{d/2}!}\bigg(\frac{\ln\ceil{d/2}}{2\ceil{d/2}}\bigg)^{\ceil{d/2}}(\alpha n)^{\ceil{d/2}} \\
    &< \frac{2^{\ceil{d/2}+1}e(d+1)(k+1)^{\floor{d/2}}}{\ceil{d/2}!}\bigg(\frac{\ln\ceil{d/2}}{\ceil{d/2}}\bigg)^{\ceil{d/2}}n^{\ceil{d/2}}.
\end{align*}

If $T_k(P) < \binom{n}{d}$, then there exists a subset of $d$ points such that any $d$-ball that contains that subset must contain more than $k$ other points of $P$ in its interior. Replacing $T_k(P)$ by our upper bound and $\binom{n}{d}$ by the well-known lower bound of $n^d / d^d$, we get
\begin{equation}
    \frac{2^{\ceil{d/2}+1}e(d+1)(k+1)^{\floor{d/2}}}{\ceil{d/2}!}\bigg(\frac{\ln\ceil{d/2}}{\ceil{d/2}}\bigg)^{\ceil{d/2}}n^{\ceil{d/2}} < \frac{n^d}{d^d}
\end{equation}
which implies
\begin{equation}
    k < \bigg(\frac{\ceil{d/2}!}{2^{\ceil{d/2}+1}e(d+1)d^d}\bigg(\frac{\ln\ceil{d/2}}{\ceil{d/2}}\bigg)^{\ceil{d/2}}\bigg)^{\frac{1}{\floor{d/2}}}n - 1.
\end{equation}
\end{proof}

Similar to \Cref{cor:convex-sets-inside-outside}, we can prove a similar result for points inside and outside any homothet of a $d$-ball with $Q$ on its boundary. Let $P$ be a set of $n$ points in $\mathbb{R}^d$. We say that a subset of $d+1$ points of $P$ form a simplex of the order-$k$ farthest-point Delaunay graph if the $d$-ball with those $d+1$ points on the boundary contains at least $n-k-d-1$ points in its interior. Let $\overline{T}_k(P)$ denote the number of $(d-1)$-simplices in the order-$k$ farthest-point Delaunay graph of $P$. We can use this to prove the existence of a set of $d$ points such that any $d$-ball with those points on the boundary must include a constant fraction of the points of $P$, and exclude a constant fraction of the points of $P$.

\begin{cor}
\label{cor:d-balls-inout}
Let $P$ be a set of $n$ points in $\mathbb{R}^d$, where $d \ge 3$. There exists a subset $Q$ of $P$ of size $d$ such that any $d$-ball with $Q$ on its boundary also contains at least
\begin{equation*}
    \bigg(\frac{\ceil{d/2}!}{2^{\ceil{d/2}+1}e(d+1)d^d}\bigg(\frac{\ln\ceil{d/2}}{\ceil{d/2}}\bigg)^{\ceil{d/2}}\bigg)^{\frac{1}{\floor{d/2}}}n - 1
\end{equation*}
other points of $P$ in its interior, and at least the same number of points are exterior to the $d$-ball (that is, contained in the interior of the complement of the $d$-ball).
\end{cor}

\begin{proof}
In the proof of \Cref{thm:d-balls}, we used the fact that the $d$-simplices of the order-$0$ Delaunay triangulation corresponds to the lower hull of the image of $P$ after lifting to the unit paraboloid in $\mathbb{R}^{d+1}$. The $d$-simplices of the order-$0$ \emph{farthest point} Delaunay triangulation corresponds to the \emph{upper hull} of the image of $P$. The union of these two is the set of facets of the convex hull of the image of $P$. Thus, the sum $T_0(P) + \overline{T}_0(P)$ is at most the number of facets of the convex hull of the image of $P$. Therefore,
\begin{equation}
    T_0(R) + \overline{T}_0(R) \le \bigg(\frac{2(d+1)}{\ceil{d/2}!}\bigg)\abs{R}^{\ceil{d/2}}
\end{equation}
and the rest of the proof follows.
\end{proof}

The proofs of \Cref{thm:d-balls} and \Cref{cor:d-balls-inout} uses a bound on the number of $(d-1)$-simplices that holds for any $d$. A better upper bound on this quantity can be computed for $\mathbb{R}^3$, which allows us to get a better constant. Note that the upper bound on $T_0(R)$ of \Cref{thm:d-balls} gives a constant of $c \approx 1/9776.59$ when $d=3$.

\begin{cor}
\label{cor:spheres}
Let $P$ be a set of $n$ points in $\mathbb{R}^3$. There exists a triple of points $\{p,q,r\}$ in $P$ such that any closed ball that contains $p$, $q$, and $r$ on its boundary also contains at least $(n-3)/6e - 1 \approx n/16.31$ other points of $P$ in its interior, and at least that many points are \emph{not} in its interior.
\end{cor}

\begin{proof}
The Upper Bound Theorem~\cite{mcmullen1970maximum,ziegler2012lectures} implies that a $4$-polytope with $n$ vertices has at most $n^2 - 3n$ triangles, so use the upper bound of
\begin{equation}
    E[T_0(R) + \overline{T}_0(R)] \le E[\abs{R}^2] - 3E[\abs{R}] = (\alpha n)^2 - (2 + \alpha)\alpha n
\end{equation}
instead in the proof of \Cref{cor:d-balls-inout}. Here, we use the fact that $E[\abs{R}^2] = (\alpha n)^2 + (1-\alpha)\alpha n$ since $\abs{R}$ is binomially distributed. The rest of the proof is identical to that of \Cref{cor:convex-sets-inside-outside}.
\end{proof}

In this section we showed that there exists a subset $Q$ of $P$ of size $d$ such that any $d$-ball that contains $Q$ must contain at least $cn$ other points in its interior, for some constant $c$. Previous work~\cite{BaranySSU89,SmorodinskySW08} for this problem in $\mathbb{R}^d$ showed the existence of a subset $Q$ of size $\floor{(d+3)/2}$ with the same property. Thus our results are not strictly comparable to previous results on this problem. However, our result for $\mathbb{R}^3$ improves the state of the art from $c = 1/60$ as proved by Barany et al.~\cite{BaranySSU89} to $c \approx 1/16.31$.

\section{Homothets of a triangle}

In Section~\ref{sec:general-convex} we considered the case where $C$ is a general convex region, proving that there exists two points $p$ and $q$ such that any homothet of $C$ that contains both $p$ and $q$ must contain at least $n/6e \approx n/16.31$ points. In this section we consider the special case where $C$ is a triangle (that is, a compact region given by the intersection of three halfplanes), and we improve the lower bound from $n/6e$ to $n/12$ for this special case. Throughout the section, we will make the general position assumption two points of $P$ lie on a line parallel to an edge of $C$.

Let $T$ be the triangle with vertices $(-\frac14,-\frac14)$, $(\frac34,-\frac14)$, and $(-\frac14,\frac34)$. All homothets of $T$ can be characterized as sets of the form
\begin{equation}
    \triangle(a,b,c) = \{(x,y) \in \mathbb{R}^2 : x \ge a, y \ge b, x + y \le c\}
\end{equation}
such that $a + b < c$. If $a + b > c$ then the set $\triangle(a,b,c)$ is empty, and if $a + b = c$ then the set $\triangle(a,b,c)$ consists only of the point $(a,b)$.

Given two points $p$ and $q$, the smallest homothet of $T$ that contains both $p$ and $q$ is
\begin{equation}
    T(p,q) = \triangle(\min\{p_x, q_x\}, \min\{p_y, q_y\}, \max\{p_x+p_y, q_x+q_y\})
\end{equation}
in the sense that if $\triangle(a,b,c)$ is a homothet of $T$ that contains $p$ and $q$, then $T(p,q)$ is a subset of $\triangle(a,b,c)$. To see this, suppose $\triangle(a,b,c)$ contains $p$ and $q$, and let $(x,y)$ be a point in $T(p,q)$. Then $p_x \ge a$ and $q_x \ge a$, and so $\min\{p_x,q_x\} \ge a$. But $x \ge \min\{p_x,q_x\}$, so $x \ge a$. Similarly we can show that $y \ge b$ and $x + y \le c$, so $(x,y)$ is in $\triangle(a,b,c)$.

\begin{lem}
\label{lem:triangles}
Let $P$ be a set of $n$ points in the plane. The average value of $\abs{T(p,q) \cap P}$ over all pairs of points $\{p,q\}$ in $P$ is at least $(n-7)/12$.
\end{lem}

\begin{proof}
For three points $p$, $q$, and $r$, at most one of the relations $r \in T(p,q)$, $p \in T(q,r)$, and $q \in T(r,p)$ hold. We say that three points $\{p,q,r\}$ is a \emph{good triple} if one of these three relations is true. Let $n(r)$ be the number of triangles $T(p,q)$ such that $r \in T(p,q)$. The number of good triples can be counted by the sum
\begin{equation}
    \sum_{r \in P} n(r).
\end{equation}
We will show that this sum is large.

First, we introduce some notation. Given a point $p$, define the three sets
\begin{equation}
    A_1(p) = \{q \in P : p_x < q_x\}, \quad
    A_2(p) = \{q \in P : p_y < q_y\}, \quad
    A_3(p) = \{q \in P : p_x + p_y > q_x + q_y\}.
\end{equation}
Let $a_i(p) = \abs{A_i(p)}$ and $a_{ij}(p) = \abs{A_i(p) \cap A_j(p)}$. Notice that $r \in A_i(p)$ if and only if $p \not\in A_i(r)$. Furthermore, $r \in T(p,q)$ if and only if, for all $i \in \{1,2,3\}$, it is not the case that both $p$ and $q$ are in $A_i(r)$.

For $i \in \{1,2,3\}$, let $E_i(r)$ be the set of unordered pairs $\{p,q\}$ such that both $p$ and $q$ are in $A_i(r)$. We have $r \in T(p,q)$ if and only if $\{p,q\} \not\in \big(E_1(r) \cup E_2(r) \cup E_3(r)\big)$, so
\begin{equation}
    n(r) = \frac{(n-1)(n-2)}{2} - \abs{E_1(r) \cup E_2(r) \cup E_3(r)}
\end{equation}

Note that $E_1(r) \cap E_2(r) \cap E_3(r)$ is empty: if $\{p,q\}$ is in both $E_1(r)$ and $E_2(r)$, then $r_x < p_x$ and $r_y < p_y$, thus $r_x + r_y < p_x + p_y$ and $p \not\in A_3(r)$. Thus by the principle of inclusion-exclusion,
\begin{equation}
    n(r) = \frac{(n-1)(n-2)}{2} - \sum_{1\le i\le3} \abs{E_i(r)} + \sum_{1\le i < j \le 3} \abs{E_i(r) \cap E_j(r)}
\end{equation}

We have
\begin{equation}
    \lvert E_i(r) \rvert = \frac{a_i(r)(a_i(r) - 1)}{2}
    \quad\text{and}\quad
    \lvert E_i(r) \cap E_j(r)\rvert = \frac{a_{ij}(r) (a_{ij}(r) - 1)}{2}
\end{equation}

All together we have
\begin{align}
\sum_{r \in P} n(r)
&= \sum_{r \in P}\frac{(n-1)(n-2)}{2} - \sum_{r \in P} \sum_{1 \le i \le 3} \lvert E_i(r) \rvert + \sum_{r \in P} \sum_{1 \le i < j \le 3} \lvert E_i(r) \cap E_j(r) \rvert \\
&= \sum_{r \in P}\frac{(n-1)(n-2)}{2} - \frac{1}{2}\sum_{r \in P} \sum_{1 \le i \le 3} a_i(r)^2 + \frac{1}{2}\sum_{r \in P} \sum_{1 \le i \le 3} a_i(r) + \frac{1}{2}\sum_{r \in P} \sum_{1 \le i < j \le 3} a_{ij}(r)^2 - \frac{1}{2}\sum_{r \in P} \sum_{1 \le i < j \le 3} a_{ij}(r)
\end{align}

We will tackle each sum separately.

Because of our general position assumption, for each $i \in \{1,2,3\}$ the multiset $\{a_i(r) : r \in P\}$ is equal to $\{0,1,\dots, n-1\}$. Hence
\begin{equation}
\sum_{r\in P} a_i(r) = \sum_{k=0}^{n-1} k = \frac{n(n-1)}{2}
\quad\text{and}\quad
\sum_{r\in P} a_i(r)^2 = \sum_{k=0}^{n-1} k^2 = \frac{n(n-1)(2n-1)}{6}
\end{equation}
and thus we have
\begin{equation}
\sum_{r \in P}\sum_{1 \le i \le 3} a_i(r) = \frac{3n(n-1)}{2}
\end{equation}
and
\begin{equation}
\sum_{r \in P}\sum_{1 \le i \le 3} a_i(r)^2 = \frac{n(n-1)(2n-1)}{2}
\end{equation}

Define $m(r,p) = \mathbf{1}[p \in A_1(r)] + \mathbf{1}[p \in A_2(r)] + \mathbf{1}[p \in A_3(r)]$. Note that $m(r,p) + m(p,r) = 3$ because $p \in A_i(r)$ if and only if $r \not\in A_i(p)$, and $m(r,p)$ cannot be $3$ since $A_1(r) \cap A_2(r) \cap A_3(r) = \varnothing$, so $m(r,p) \in \{1,2\}$. Exactly one of $m(r,p)$ and $m(p,r)$ is $2$, so if $M_2$ is the number of pairs such that $m(p,r) = 2$ then $M_2 = n(n-1)/2$.

Finally, note that
\begin{equation}
M_2 = \sum_{r \in P} \sum_{1 \le i < j \le 3} a_{ij}(r)
\end{equation}
since if $m(r,p) = 2$, then $p$ will be in exactly one set of the form $A_i(r) \cap A_j(r)$. Thus
\begin{equation}
\sum_{r \in P} \sum_{1 \le i < j \le 3} a_{ij}(r) = \frac{n(n-1)}{2}.
\end{equation}

Finally we will use the Cauchy-Schwarz inequality applied do the $3n$ numbers $a_{ij}(r)^2$:
\begin{equation}
\sum_{r \in P} \sum_{1 \le i < j \le 3} a_{ij}(r)^2
\ge \frac{1}{3n}\bigg(\sum_{r \in P} \sum_{1 \le i < j \le 3} a_{ij}(r)\bigg)^2
= \frac{1}{3n}\bigg(\frac{n(n-1)}{2}\bigg)^2 = \frac{n(n-1)^2}{12}
\end{equation}

Finally we can put it all together
\begin{align}
\sum_{r \in P} n(r)
&= \sum_{r \in P}\frac{(n-1)(n-2)}{2} - \frac{1}{2}\sum_{r \in P} \sum_{1 \le i \le 3} a_i(r)^2 + \frac{1}{2}\sum_{r \in P} \sum_{1 \le i \le 3} a_i(r) + \frac{1}{2}\sum_{r \in P} \sum_{1 \le i < j \le 3} a_{ij}(r)^2 - \frac{1}{2}\sum_{r \in P} \sum_{1 \le i < j \le 3} a_{ij}(r) \\
&\ge \frac{n(n-1)(n-2)}{2} - \frac{1}{2}\bigg(\frac{n(n-1)(2n-1)}{2}\bigg) + \frac{1}{2}\bigg(\frac{3n(n-1)}{2}\bigg) + \frac{1}{2}\bigg(\frac{n(n-1)^2}{12}\bigg) - \frac{1}{2}\bigg(\frac{n(n-1)}{2}\bigg) \\
&= \frac{n(n-1)(n-7)}{24}
\end{align}

We can write the number of good triples in another way.  The number of good triples is the sum over all triangles $T(p,q)$ of the number of points of $P$ contained in that triangle. That is,
\begin{equation}
\sum_{r \in P} n(r) = \frac{1}{2}\sum_{p \ne q} (\lvert T(p,q) \cap P\rvert - 2)
\end{equation}
the factor of $1/2$ comes from the fact that $T(p,q) = T(q,p)$ so the sum counts each triangle twice.

The average number of points in $T(p,q)$ over all pairs is
\begin{equation}
\frac{1}{n(n-1)}\sum_{p \ne q} (\lvert T(p,q) \cap P\rvert - 2).
\end{equation}
Therefore, the average number of points in $T(p,q)$ is at least
\begin{equation}
\frac{1}{n(n-1)}\sum_{p \ne q} (\lvert T(p,q) \cap P\rvert - 2) = \frac{2}{n(n-1)}\sum_{r \in P} n(r) \ge \frac{n-7}{12}.
\end{equation}
In particular, there must exist a pair $\{p,q\}$ such that $T(p,q)$ contains at least $(n-7)/12$ points of $P$.
\end{proof}

\begin{thm}
\label{thm:triangles}
Let $T$ be a triangle in the plane that contains the origin in its interior. Let $P$ be a set of $n$ points in the plane such that no two points lie on a line parallel to an edge of $T$. Then there exists a pair of points $p$ and $q$ such that any homothet of $T$ that contains both $p$ and $q$ must contain at least $(n-7)/12$ other points of $P$.
\end{thm}

\begin{proof}
There is a unique affine transformation that maps the three vertices of $T$ to $(-\frac14,-\frac14)$, $(\frac34,-\frac14)$, and $(-\frac14,\frac34)$. Since affine transformations are invertible, preserve containment, and send homothets of $T$ to homothets of the image of $T$, we can assume without loss of generality that the vertices of $T$ are $(-\frac14,-\frac14)$, $(\frac34,-\frac14)$, and $(-\frac14,\frac34)$.

By \Cref{lem:triangles} the average value of $\abs{T(p,q) \cap P}$ is at least $(n-7)/12$. In particular, there must exist two points $p$ and $q$ such that $T(p,q)$ contains at least $(n-7)/12$ points of $P$. Since any homothet of $T$ that contains both $p$ and $q$ must contain $T(p,q)$, the result follows.
\end{proof}

As before, we can also prove that there exist two points such that any homothet of a given triangle with those two points on its boundary must contain a constant fraction of the points, while also leaving out some constant fraction of the points.

\begin{thm}
\label{thm:triangles-inout}
Let $T$ be a triangle in the plane that contains the origin in its interior. Let $P$ be a set of $n$ points in the plane such that no two points of $P$ lie on a line parallel to an edge of $T$. Then there exists a pair of points $\{p,q\}$ in $P$ such that any homothet of $T$ that has both $p$ and $q$ on its boundary must contain at least $(n-7)/15$, and at most $14(n+7)/15$, points of $P$.
\end{thm}

\begin{proof}
As in the proof of \Cref{thm:triangles}, we assume without loss of generality that the vertices of $T$ have coordinates $(-\frac14,-\frac14)$, $(\frac34,-\frac14)$, and $(-\frac14,\frac34)$.

Let $k$ be a positive integer such that $3k < n$. Partition $P$ into the following four sets:
\begin{itemize}
    \item Let $P_1$ be the $k$ elements of $P$ with the smallest $x$-coordinate.
    \item Let $P_2$ be the $k$ elements of $P \setminus P_1$ with the smallest $y$-coordinate.
    \item Let $P_3$ be the $k$ elements of $P \setminus (P_1 \cup P_2)$ with the largest sum of coordinates.
    \item Let $P_4 = P \setminus (P_1 \cup P_2 \cup P_3)$.
\end{itemize}

By \Cref{lem:triangles}, there exist two points in $P_4$ such that any homothet of $T$ that contains $p$ and $q$ must contain at least $(n-3k-7)/12$ points of $P_4$. In particular, this includes all homothets of $T$ with $p$ and $q$ on its boundary.

Because of our general position assumption, exactly one of $p$ and $q$ lies on a vertex of $T(p,q)$, and the other point lies on the opposite edge of $T(p,q)$. Suppose that $p$ lies on the left edge of $T(p,q)$, and $q$ lies on the rightmost vertex. Any homothet of $T$ with $p$ and $q$ on its boundary must have $p$ on its left edge. Thus, since $p$ is in $P_4$, and every point of $P_1$ has a smaller $x$-coordinate than $p$, any homothet of $T$ with $p$ and $q$ on its boundary contains at most $n-k$ points.

Similar arguments apply when $p$ is on the other two edges of $T(p,q)$. In all cases, at least one of the sets $P_1$, $P_2$, or $P_3$ must be disjoint from $T$. So every homothet of $T$ with $p$ and $q$ on its boundary contains at most $n-k$ points. Setting $k=(n-7)/15$ completes the proof.
\end{proof}

\section{Homothets of a parallelepiped in three dimensions}

We now consider a different special case. Let $C$ be a parallelepiped in $\mathbb{R}^3$, meaning a polyhedron with six paralellogram faces. A parallelepiped is the result of applying an affine transformation to a cube. We will show that there exist two points $p$ and $q$ such that any homothet of $C$ that contains both $p$ and $q$ contains at least (approximately) $n/768$ points of $P$. Prior to this paper, the best lower bound was $n/2048$ by Ishigami~\cite{Ishigami95}. \citet{kiem2023four} proved the following result which we will need in our proof.

\begin{lem}
\label{lem:4ct}
Let $n\in\mathbb{N}$ and let $\chi:E(K_n)\to\{1,2,3,4\}$ be any $4$-edge-coloring of $K_n$.
Let $M(\chi)$ denote the number of monochromatic triangles in $\chi$.
Then for every $\varepsilon>0$ there exists $n_0=n_0(\varepsilon)$ such that for all $n\ge n_0$,
\[
M(\chi)\ \ge\ \left(\frac{1}{256}-\varepsilon\right)\binom{n}{3}.
\]
\end{lem}

Let $B(p,q)$ be the smallest axis-aligned box that contains $p = (x_p, y_p, z_p)$ and $q = (x_q, y_q, z_q)$. More precisely,
\begin{equation}
    B(p,q) = [\min\{x_p, x_q\}, \max\{x_p, x_q\}] \times [\min\{y_p, y_q\}, \max\{y_p, y_q\}] \times [\min\{z_p, z_q\}, \max\{z_p, z_q\}].
\end{equation}
Notice that if $C$ is any rectangular cuboid, then a homothet of $C$ that contains two points $p$ and $q$ must contain $B(p,q)$. Similar to the previous section, we will first prove that the average number of points of $P$ contained in $B(p,q)$ is large, then the result will follow from applying an affine transformation.

\begin{lem}
\label{lem:boxes}
Let $P$ be a set of $n$ points in $\mathbb{R}^3$ such that no two points share an $x$-coordinate, $y$-coordinate,or $z$-coordinate. The average value of $\abs{B(p,q) \cap P}$ over all pairs of points $\{p,q\}$ in $P$ is at least $(1/768 - o(1))n$.
\end{lem}

\begin{proof}
Any axis-parallel box containing $p$ and $q$ also contains $B(p,q)$, so it suffices to find $p,q$ with
$|B(p,q)\cap P|$ large.

Let $P = \{p_1, p_2, \dots, p_n\}$, where $p_i = (x_i, y_i, z_i)$. We also assume that the points are numbered so that $x_1 < x_2 < x_3 < \cdots < x_n$.

Consider the complete graph with vertex set $\{1, 2, \dots, n\}$. We will colour the edges of this graph so that the subgraph induced by $\{i,j,k\}$, where $i < j < k$, is monochromatic if and only if $p_j$ is contained in $B(p_i, p_k)$. Consider an edge $\{i,j\}$ with $i < j$. Colour the edge by
\begin{equation}
    c(i,j) = (\mathbf{1}[y_i > y_j], \mathbf{1}[z_i > z_j]) \in \{0,1\}^2,
\end{equation}
where $\mathbf{1}[\phi] = 1$ if $\phi$ is true, and $\mathbf{1}[\phi] = 0$ if $\phi$ is false. This is a $4$-edge-colouring of the complete graph.

For $i < j < k$, the triangle $\{i,j,k\}$ is monochromatic if and only if $(y_i, y_j, y_k)$ and $(z_i, z_j, z_k)$ are monotone sequences. That is, either $y_i < y_j < y_k$ or $y_i > y_j > y_k$, and similarly for the $z$-coordinates. It follows that $p_j$ is in $B(p_i, p_k)$ if and only if the triangle $\{i,j,k\}$ is monochromatic in this edge colouring.

Let $M$ be the total number of monochromatic triangles in this edge colouring.
We have $\sum_{i < k} \big(\lvert B(p_i, p_k) \rvert - 2\big) = M$, since every monochromatic triangle corresponds to exactly one triple of points $\{p_i, p_j, p_k\}$ such that $p_j \in B(p_i, p_k)$.
By \Cref{lem:4ct}, we know that $M \ge (1/256 - o(1))\binom{n}{3}$. 
Therefore if we choose a pair of points uniformly at random, the expected value of $\lvert B(p_i, p_k) \rvert - 2$ is $M / \binom{n}{2} \ge (1/768 - o(1))n$.
Since the expected value is at least (approximately) $n/768$, there must exist a pair of points $p_i, p_k$ such that $B(p_i, p_k)$ contains at least $n/768$ points.
\end{proof}

\begin{thm}
\label{thm:boxes}
Let $B$ be a parallelepiped in $\mathbb{R}^3$ that contains the origin in its interior. Let $P$ be a set of $n$ points in $\mathbb{R}^3$ such that no two points lie on a line parallel to a face of $B$. Then there exists a pair of points $p$ and $q$ such that any homothet of $B$ that contains both $p$ and $q$ must contain at least $(1/768 - o(1))n$ other points of $P$.
\end{thm}

\begin{proof}
There exists an affine transformation that maps $B$ to a cuboid $B'$ whose opposite faces are parallel, and any homothet of the $B'$ that contains two given points $p$ and $q$ must also contain $B(p,q)$. The result follows because affine transformations are invertible and preserve containment.
\end{proof}

We note that the result of \citet{kiem2023four} holds asymptotically, so our result is only true for sufficiently large sets of points. Additionally, this approach will prove a lower bound for parallelepipeds in any dimension. In $d$ dimensions, we will get an edge-colouring of the complete graph $K_n$ that uses $2^{d-1}$ colours. It is known that in any edge-colouring of $K_n$ with two different colours, asymptotically a fraction of at least $1/4$ of all triangles are monochromatic \cite{goodman1959sets}. This would imply a lower bound of $n/12$ for homothets of a parallelogram in the plane, however as we noted earlier the best-known lower bound is $n/5$.

\section{Homothets of a triangle with rotation}

We now consider generalizing the problem to the case where $C$ is a triangle in the plane, and in addition to homothets (uniform scaling and translations), we allow $C$ to be rotated by $\pi$ radians. Equivalently, we consider sets of the form $\lambda C + x$, where $\lambda$ can be any real number (as opposed to requiring that $\lambda$ be positive). These are equivalent since the set $-C = \{p \in \mathbb{R}^2 : -p \in C\}$ is the result of rotating $C$ by $\pi$ radians about the origin. We prove a lower bound for this case using the results of the previous section on parallelepipeds in $\mathbb{R}^3$.

\begin{figure}
    \centering
    \includegraphics{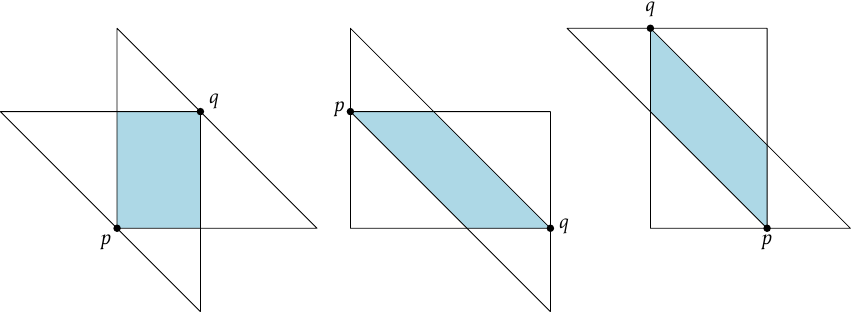}
    \caption{Depending on how the points $p$ and $q$ are oriented, the region $T(p,q) \cap T'(p,q)$ (shaded) can have different shapes.}
    \label{fig:two-triangles}
\end{figure}

Let $T$ be the triangle with vertices $(-\frac14,-\frac14)$, $(\frac34,-\frac14)$, and $(-\frac14,\frac34)$. Recall that we defined $T(p,q)$ to be the smallest homothet of $T$ that contains $p$ and $q$. Let $T'$ be the result of rotating $T$ by an angle of $\pi$ radians. Define $T'(p,q)$ to be the smallest homothet of $T'$ that contains $p$ and $q$. The smallest homothet of $T'$ that contains two points $p$ and $q$ is the set
\begin{equation}
    T'(p,q) = \{(x,y) \in \mathbb{R}^2 : x \le \max\{p_x, q_x\}, y \le \max\{p_y, q_y\}, x + y \ge \min \{p_x + p_y, q_x + q_y\}\}.
\end{equation}
The shape of the region $T(p,q) \cap T'(p,q)$ depends on the orientation of the points $p$ and $q$, see Figure~\ref{fig:two-triangles}.

\begin{lem}
Let $P$ be a set of $n$ points in the plane such that no two points share an $x$-coordinate, $y$-coordinate, or lie on a line with slope $-1$. Then the average value of $\abs{T(p,q) \cap T'(p,q) \cap P}$ over all pairs of points $p$ and $q$ in $P$ is at least $(1/768 - o(1))n$.
\end{lem}

\begin{proof}
Consider the function $f : \mathbb{R}^2 \to \mathbb{R}^3$ that sends the point $(x,y)$ to $(x,y,x+y)$. For two points $p$ and $q$ in the plane, the box $B(f(p), f(q))$ in $\mathbb{R}^3$ is defined by the six inequalities
\begin{equation}
\begin{split}
    \min\{p_x, q_x\} &\le \mathord{\makebox[\widthof{$x+y$}]{$x$}} \le \max\{p_x, q_x\}, \\
    \min\{p_y, q_y\} &\le \mathord{\makebox[\widthof{$x+y$}]{$y$}} \le \max\{p_y, q_y\}, \\
    \min\{p_x+p_y, q_x+q_y\} &\le x + y \le \max\{p_x+p_y, q_x+q_y\},
\end{split}
\end{equation}
which are exactly the six inequalities that imply $(x,y) \in T(p,q) \cap T'(p,q)$. Therefore, the number of points of $P$ contains in $T(p,q) \cap T'(p,q)$ is equal to the number of points of $f(P)$ contained in $B(f(p), f(q))$. By~\Cref{lem:boxes}, we know that the average value of $\abs{f(P) \cap B(f(p), f(q))}$ is at least $(1/768 - o(1))n$, so the average value of $\abs{T(p,q) \cap T'(p,q) \cap P}$ has the same lower bound.
\end{proof}

As in the previous two sections, the main result of this section follows almost directly from the lemma.

\begin{thm}
Let $T$ be a triangle in the plane that contains the origin in its interior. Let $T'$ be the result of rotating $T$ by an angle of $\pi$ radians about the origin. Let $P$ be a set of $n$ points in the plane such that no two points lie on a line parallel to an edge of $T$. Then there exists a pair of points $p$ and $q$ such that any homothet of either $T$ or $T'$ that contains both $p$ and $q$ must contain at least $(1/768 - o(1))n$ other points of $P$.
\end{thm}

\begin{proof}
As in the proof of \Cref{thm:triangles}, we can assume without loss of generality that the three vertices of $T$ are $(-\frac14,-\frac14)$, $(\frac34,-\frac14)$, and $(-\frac14,\frac34)$.

Since the average value of $\abs{T(p,q) \cap T'(p,q) \cap P}$ is at least $(1/768 - o(1))n$, there must be a pair of points $p$ and $q$ of $P$ such that $\abs{T(p,q) \cap T'(p,q) \cap P} \ge (1/768 - o(1))n$. Every homothet of $T$ that contains $p$ and $q$ contains $T(p,q)$, and every homothet of $T'$ that contains $p$ and $q$ contains $T'(p,q)$.
\end{proof}

\section{Algorithms}

In this section we consider the algorithmic problem of \emph{finding} two points $p$ and $q$ such that any homothet of $C$ that contains $p$ and $q$ contains $cn$ points of $P$, for some constant $c$. Let $N_C(P,p,q)$ be the minimum number of points of $P$ contained in any homothet of $C$ that contains $p$ and $q$. In previous sections, we gave several proofs of the existence of a pair of points such that $N_C(p,q) \ge cn$, each with different assumptions about $C$.

\subsection{Strictly convex sets}

We first consider the case where $C$ is a strictly convex region. Recall that we defined $\lambda_k$ to be the number of connected components summed over each bisector of the union of the order-$1$ through order-$k$ Voronoi diagrams. This was an upper bound on the number of bisectors that did not contain a segment of weight $k$. If $\lambda_k < (1-\alpha)\binom{n}{2}$ for $0 < \alpha < 1$, then at least $\alpha\binom{n}{2}$ bisectors have no segments of weight less than $k$. Therefore choosing a pair $\{p,q\}$ uniformly at random, we will have $N_C(p,q) \ge k$ with probability $\alpha$. What remains to be seen is how to compute $N_C(p,q)$ for a given convex distance function. \citet{BoseET25} presented an algorithm for computing $N_C(p,q)$ for the case where $C$ is a disk in the plane. Their algorithm computes the weight of every segment along $b_C(p,q)$.

Ma~\cite{Ma00} proved that, assuming $C$ is strictly convex, the bisector of three points $b_C(p,q,r)$, which is the set of points $s$ such that $d_C(p,s)=d_C(q,s)=d_C(r,s)$, is either empty if $p$, $q$, and $r$ are collinear, or consists of a single point. They also give an algorithm that uses prune-and-search to construct a sequence of points that approaches $b_C(p,q,r)$. This algorithm relies on being able to construct the intersection point of a ray with the boundary of $C$. Because we make no assumption about the representation of the boundary, we will simply assume that this operation can be carried out in time $T$. For a case such as where $C$ is a disk, this is a constant time operation. For more complicated boundaries, $T$ may depend on some other parameter. For example, Bose et al.~\cite{BoseET25} showed how to implement this algorithm for sets of points in a simple polygon, where computing the centre of the geodesic disk through three given points requires time proportional to the logarithm of the number of vertices of the enclosing polygon~\cite{DBLP:journals/dcg/OhA20}.

The following lemma about bisectors of convex distance functions will help us in designing efficient algorithms for our problem.

\begin{lem}
\label{lem:monotone-bisector}
Let $C$ be a strictly compact, convex set that contains the origin in its interior, and let $p$ and $q$ be two points. The bisector $b_C(p,q)$ is a monotone curve with respect to the Euclidean bisector of $p$ and $q$.
\end{lem}

\begin{figure}
    \centering
    \includegraphics{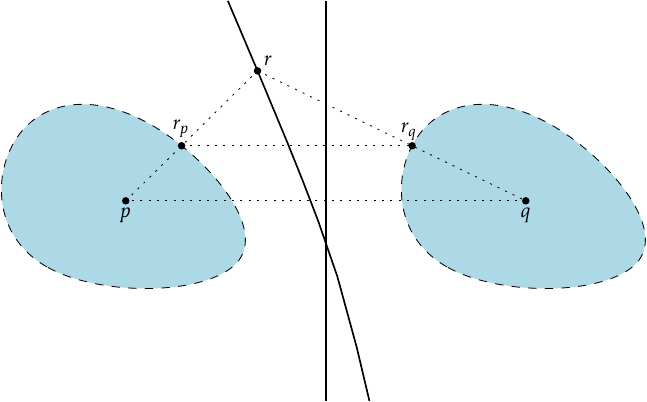}
    \caption{Two points $p$ and $q$, and a point $r$ on $b_C(p,q)$. The segments $pq$ and $r_pr_q$ are parallel. The Euclidean bisector of $p$ and $q$ is also shown.}
    \label{fig:bisector}
\end{figure}

\begin{proof}
Let $r$ be a point on $b_C(p,q)$. Then let $r_p$ be the intersection of $C + p$ with the ray originating at $p$ that passes through $r$, and let $r_q$ be defined similarly for $q$. Ma~\cite[Lemma~2.1.1.1]{Ma00} showed that $pq$ and $r_pr_q$ are parallel, and that the mapping from $r$ to $r_p$ is a homeomorphism. See Figure~\ref{fig:bisector}. Thus the angle $\angle qpr$ is monotone as $r$ moves along $b_C(p,q)$. If there were two points $r$ and $r'$ on $b_C(p,q)$ such that $rr'$ and $pq$ were parallel, then one of $r_pr_q$ or $r'_pr'_q$ would not be parallel with $pq$.
\end{proof}

We are now ready to show how to compute $N_C(p,q)$ given $p$ and $q$.

\begin{thm}
\label{thm:alg-bisector-segments}
Let $P$ be a set of $n$ points in the plane, and let $C$ be a compact, strictly convex subset of the plane that contains the origin in its interior. There exists an algorithm to compute $N_C(P,p,q)$ given two points $p$ and $q$ in $P$ with time complexity $O(Tn + n \log n)$, where $T$ is the time complexity of finding the centre of the homothet of $C$ with three given points on its boundary.
\end{thm}

\begin{proof}
Recall that we can partition the bisector $b_{-C}(p,q)$ into $n-1$ segments. The $n-2$ boundaries of these segments are the centres of the $n-2$ homothets of $C$ with $p$, $q$, and a third point of $P$ on their boundary. We will compute the weight of every segment on $b_{-C}(p,q)$. Since the weight of two adjacent segments differs by one, we will first find the weight of one of the unbounded segments, then sort the segments along $b_{-C}(p,q)$.

The weight of one of the unbounded segments of $b_{-C}(p,q)$ is equal to the number of points on one side of the line through $p$ and $q$. Therefore, in $O(n)$ time we can compute the weight of one of these unbounded segments. To find the order of the segments along $b_{-C}(p,q)$, we first find the centres of the $n-2$ homothets of $C$ that contain $p$, $q$, and a third point of $P$ on its boundary. These centres can be found in $O(Tn)$ time, where $T$ is the time complexity of computing the centre of the homothet of $C$ with three given points on its boundary. Then, because of \Cref{lem:monotone-bisector}, we can project all of these centres onto the Euclidean bisector of $p$ and $q$, and sort them along that line. Sorting these points takes $O(n \log n)$ time.

Given the weight of one segment $s$ on $b_{-C}(p,q)$, computing the weight of the next segment $s'$ is simple. Let $r$ be the third point of $P$ such that $p$, $q$, and $r$ are on the boundary of the homothet centred at the intersection of $s$ and $s'$. Because of \Cref{lem:monotone-bisector}, we can say without loss of generality that $s'$ is to the right of $s$. Then, if $r$ is to the right of the line through $p$ and $q$, then the weight of $s'$ is one more than the weight of $s$. Otherwise, the weight of $s'$ is one less than the weight of $s$. Thus we can compute in constant time the weight of $s'$ given the weight of $s$. This implies that we can compute the weight of every segment on $b_{-C}(p,q)$ in $O(n)$ time.
\end{proof}

Finding the homothet that has three given points on its boundary is the only operation we require where the shape of $C$ actually matters. If $C$ is a disk, then this is trivial to do in constant time. Ma~\cite{Ma00} gives an algorithm using prune-and-search that produces a point which approaches this centre, although it relies on being able to compute the intersection of a ray with the boundary of $C$. The exact time it takes to compute this intersection depends on the complexity of $C$. They also give an algorithm for the case where $C$ is a simple polygon with $k$ sides that takes $O(\log k + m)$ time, where $m$ is the number of vertices on the bisector of the three points (since simple polygons are not strictly convex, the bisector of three points may contain more than a single point). However, to be as general as possible we only assume that it is possible to find this centre, without making any assumption about how long this takes.

\begin{thm}
\label{thm:alg-strictly-convex}
Let $0 < \alpha < 1$ be a constant, let $P$ be a set of $n$ points in the plane, and let $C$ be a compact, strictly convex subset of the plane that contains the origin in its interior. There exists an algorithm to find two points $p$ and $q$ in $P$ such that $N_C(P,p,q) \ge (3 - \sqrt{4 + 5\alpha})n/5$ with expected time complexity $O(Tn + n \log n)$, where $T$ is the time complexity of finding the centre of the homothet of $C$ with three given points on its boundary.
\end{thm}

\begin{proof}
If $\lambda_k < (1 - \alpha)\binom{n}{2}$, then at least $\alpha\binom{n}{2}$ bisectors have no segments of weight $k$ or less. Therefore if we choose a pair of points $\{p,q\}$ of $P$ uniformly at random, we have $N_C(P,p,q) \ge k$ with probability $\alpha$. If we choose pairs until the inequality holds, the expected number of pairs we need to choose is $1/\alpha$. By \Cref{thm:alg-bisector-segments}, we can compute $N_C(P,p,q)$ in time $O(Tn + n \log n)$.

From the proof of \Cref{thm:strictly-convex-sets}, we know that $\lambda_k \le 3kn - \frac52k^2 - \frac52k$. Therefore we need $3kn - \frac52k^2 - \frac52k < (1 - \alpha)\binom{n}{2}$. Solving this inequality for $k$ gives $k \ge (3 - \sqrt{4 + 5\alpha})n/5$.
\end{proof}

We can modify the algorithm of \Cref{thm:alg-bisector-segments} slightly to also compute the maximum number of points contained in any homothet of $C$ with two given points on the boundary. Since it computes the weight of every segment of $b_{-C}(p,q)$, we simply find the maximum weight in addition to the minimum weight. Thus we can modify the algorithm of \Cref{thm:alg-strictly-convex} to find a pair of points $p$ and $q$ such that any homothet of $C$ with $p$ and $q$ on its boundary contains at least $cn$, and at most $(1-c)n$, points of $P$.

\begin{cor}
\label{cor:alg-strictly-convex-inside-outside}
Let $0 < \alpha < 1$ be a constant, let $P$ be a set of $n$ points in the plane, and let $C$ be a compact, strictly convex subset of the plane that contains the origin in its interior. There exists an algorithm to find two points $p$ and $q$ in $P$ such that every homothet of $C$ that has $p$ and $q$ on its boundary contains at least $(7 - \sqrt{37 + 12\alpha})n/12$, and at most $(1 - (7 - \sqrt{37 + 12\alpha})/12)n$, other points of $P$, with expected time complexity $O(Tn + n \log n)$, where $T$ is the time complexity of finding the centre of the homothet of $C$ with three given points on its boundary.
\end{cor}

\begin{proof}
The proof is very similar to that of \Cref{thm:alg-strictly-convex}, however we now use the upper bound of $\lambda_k + \overline\lambda_{n-k} \le 7kn - 6k^2 - 6k$ from \Cref{cor:convex-sets-inside-outside} to ensure that there exists a bisector $b_{-C}(p,q)$ such that every segment has weight between $k$ and $n-k$. We have $7kn - 6k^2 - 6k < (1 - \alpha)\binom{n}{2}$ as long as $k < (7 - \sqrt{37 + 12\alpha})n/12$.
\end{proof}

\subsection{Special cases}

The special cases where $C$ is a triangle (with or without rotation by $\pi$ radians) in the plane or parallelepiped in $\mathbb{R}^3$ are simpler. In both of these cases we have a ``minimal homothet'' that contains $p$ and $q$, in the sense that any homothet of $C$ that contains $p$ and $q$ must contain this minimal homothet. Thus our problem reduces to finding a pair of points such that the minimal homothet of those two points contains at least $cn$ points. In all three special cases, it is straightforward to compute the number of points of $P$ contained in the minimal homothet, and we have a proof that the average value of $N_C(p,q)$ is bounded below by $cn$ for some constant $c$. This second fact, coupled with the fact that $N_C(p,q) \le n$, allows us to prove that there must exist many pairs such that $N_C(p,q)$ is at least $c'n$ for some constant $c' < c$. This argument is formalized by the following lemma.

\begin{lem}
\label{lem:random-variable}
Let $X : S \to \{0,1,2,\dots,k\}$ be a random variable. Suppose the probability of any outcome in $S$ is equally likely and $E[X] \ge \alpha k$ for some $0 < \alpha < 1$. Then $\Pr(X \ge \beta k) \ge (\alpha-\beta)/(1-\beta)$ for any $0 < \beta < \alpha$.
\end{lem}

\begin{proof}
Let $A$ be the set of outcomes such that $X < \beta k$ and let $B$ be the set of outcomes such that $X \ge \beta k$. Then we have
\begin{align*}
E[X] &\le \frac{\lvert A\rvert}{\lvert S\rvert}\beta k + \frac{\lvert B\rvert}{\lvert S\rvert}k \\
&= \frac{\lvert S\rvert - \lvert B\rvert}{\lvert S\rvert}\beta k + \frac{\lvert B\rvert}{\lvert S\rvert}k \\
&= \beta k + \frac{\lvert B\rvert}{\lvert S\rvert}(1-\beta)k.
\end{align*}
Rearranging this and using the fact that $E[X] \ge \alpha k$ gives
\begin{equation*}
    \abs{B} \ge \frac{\alpha-\beta}{1-\beta}\abs{S}.
\end{equation*}
The result follows because $\Pr(X \ge \beta k) = \abs{B} / \abs{S}$.
\end{proof}

With~\Cref{lem:random-variable} in hand, we are able to give efficient algorithms for all of the special cases we considered in previous sections. All three algorithms are similar: choose a pair of points at random and count how many points are in the minimal homothet defined by the pair. Repeat this until a pair of points such that the minimal homothet contains many points is found.

\begin{thm}
\label{thm:alg-triangles}
Let $P$ be a set of $n$ points in the plane, let $T$ be a triangle in the plane that contains the origin in its interior, and let $0 < \gamma < 1/12$. There exists an algorithm to find two points $p$ and $q$ in $P$ such that any homothet of $T$ that contains both $p$ and $q$ must contain at least $\big(1 - \frac{11}{12(1-\gamma)}\big)n$ other points of $P$, that has an expected time complexity of $O(\gamma^{-1}n)$.
\end{thm}

\begin{proof}
As usual, we assume without loss of generality that the vertices of $T$ are $(-\frac14,-\frac14)$, $(\frac34,-\frac14)$, and $(-\frac14,\frac34)$. In $O(n)$ time, we can compute the value of $\abs{T(p,q) \cap P}$ for any two points $p$ and $q$ in $P$ by iterating over the points of $P \setminus \{p,q\}$ and checking if each of them is in $T(p,q)$ or not.

\Cref{lem:triangles} tells us that the average value of $\abs{T(p,q) \cap P}$ is at least $(n-7)/12$. Applying~\Cref{lem:random-variable} with $X$ being the set of all pairs of points of $P$, $k=n$, $\alpha=1/12$, and $\beta=(\alpha-\gamma)/(1-\gamma)$ implies that, for any $0 < \gamma < 1/12$, if we choose a pair $\{p,q\}$ uniformly at random, then with probability at least $\gamma$, we will have $\abs{T(p,q) \cap P} \ge \big(1 - \frac{11}{12(1-\gamma)}\big)n$. The expected number of pairs that need to be chosen before this condition is met is $\gamma^{-1}$.
\end{proof}

The proofs of the next three theorems are nearly identical to the proof of \Cref{thm:alg-triangles}, in that they are applications of \Cref{lem:random-variable}, and so are omitted. \Cref{thm:alg-triangles-inout} is slightly different in that the partition of $P$ from the proof of \Cref{thm:triangles-inout} must be computed first, but this can be done in linear time.

\begin{thm}
\label{thm:alg-triangles-inout}
Let $P$ be a set of $n$ points in the plane, let $T$ be a triangle in the plane that contains the origin in its interior, and let $0 < \gamma < 1/15$. There exists an algorithm to find two points $p$ and $q$ in $P$ such that any homothet of $T$ that contains both $p$ and $q$ must contain at least $\big(1 - \frac{14}{15(1-\gamma)}\big)n$ other points of $P$, that has an expected time complexity of $O(\gamma^{-1}n)$.
\end{thm}

\begin{thm}
Let $P$ be a set of $n$ points in $\mathbb{R}^3$, let $B$ be a parallelepiped in $\mathbb{R}^3$ that contains the origin in its interior, and let $0 < \gamma < 1/768$. There exists an algorithm to find two points $p$ and $q$ in $P$ such that any homothet of $B$ that contains both $p$ and $q$ must contain at least $(1 - \frac{767}{768(1-\gamma)})n$ other points of $P$, that has an expected time complexity of $O(\gamma^{-1}n)$.
\end{thm}

\begin{thm}
Let $P$ be a set of $n$ points in the plane, let $T$ be a triangle in the plane that contains the origin in its interior, and let $0 < \gamma < 1/768$. Let $T'$ be the result of rotating $T$ by $\pi$ radians about the origin. There exists an algorithm to find two points $p$ and $q$ in $P$ such that any homothet of either $T$ or $T'$ that contains both $p$ and $q$ must contain at least $(1 - \frac{767}{768(1-\gamma)})n$ other points of $P$, that has an expected time complexity of $O(\gamma^{-1}n)$.
\end{thm}

\section{Conclusion}

In this paper, we studied a generalization of a problem of Neuman-Lara and Urrutia. We showed that, for any convex set $C$ and any set $P$ of $n$ points in the plane, there exist two points $p$ and $q$ of $P$ such that any homothet of $C$ that contains both $p$ and $q$ contains at least $cn$ other points of $P$, for some constant $c$. We gave several proofs that achieve different constants, each with different assumptions about the set $C$. We also considered the computational aspects of finding these pairs of points.

It is interesting to note that, in the case where $C$ is a disk, the best-known lower bound on $c$ is approximately $1/4.7$~\cite{EdelsbrunnerHSX89}, while there is an upper bound of $1/4$~\cite{HaywardRW89}. On the other hand, when $C$ is a square, the constant $c$ is known to be exactly $1/5$~\cite{Grolmusz92}. Thus the constant $c$ should depend somehow on the shape of the convex region $C$. Our most general proof only assumes that $C$ is compact and convex. We showed how the ideas of this general proof can be used to prove a lower bound for a version of the problem in $\mathbb{R}^d$, in the special case where $C$ is a $d$-ball.

In addition, we proved a lower bound in the case where $C$ is a triangle, and we are allowed to rotate $C$ by $\pi$ radians in addition to scaling and translating. One avenue for future work is to consider what happens when we are allowed to rotate $C$ by any angle.

There are also several more variants of this problem that have been studied in the literature that could also be studied in the more general setting. For example, if $P$ is required to be in convex position or if the set $P$ is bichromatic, and we search for a bichromatic pair of points.

\bibliographystyle{plainurlnat}
\bibliography{ref}

\end{document}